# CrSe₂ and CrTe₂ Monolayers as Efficient Air Pollutants Nanosensors

Hakkim Vovusha [a], Puspamitra Panigrahi [b], Yash Pal [b], Muhammad J. A. Shiddiky[c], Massimiliano Di Ventra[d], Hoonkyung Lee[*a], Tanveer Hussain[*e]

[a] Advanced Materials Program, Department of Physics, Konkuk University, Seoul 05029, Republic of Korea

[b] Centre for Clean Energy and Nano Convergence, Hindustan Institute of Technology and Science, Chennai, 603103, India

[c] Rural Health Research Institute (RHRI), Charles Sturt University, Orange, NSW 2800, Australia

[d]Department of Physics, University of California, San Diego, La Jolla, California 92093, USA

[e]School of Science and Technology, University of New England, Armidale, New South Wales 2351, Australia

Corresponding authors: hkiee3@konkuk.ac.kr, tanveer.hussain@une.edu.au

## Abstract

Nanosensors are critical in environmental monitoring, industrial safety, and public health by detecting specific hazardous gases like CO, NO, SO₂, and CH₄ at trace levels. This study uses density functional theory (DFT) calculations to examine the gas-sensing capabilities of chromium diselenide (CrSe₂) and chromium ditelluride (CrTe₂) monolayers through their structural and electronic responses to gas adsorption. Adsorption energy ($E_{ads}$) analysis shows that Te vacancy-induced CrTe₂ (V$_{Te}$-CrTe₂) exhibits the strongest binding with energies of -1.52, -1.79, and -1.61 eV for CO, NO, and SO₂, respectively. Similarly, CrSe₂ has $E_{ads}$ values of -1.13, -1.17, -0.90, and -1.12 eV for CO, NO, SO₂, and CH₄, respectively, indicating suitability for reversible sensing. This study also investigates how substitutional doping of Ge, Sb, and Sn influences the sensing mechanism of CrSe₂ and CrTe₂ monolayers. Density of states (DOS) analysis highlights notable electronic changes around the Fermi level, especially in V$_{Te}$-CrTe₂ and Sb/Sn-doped CrTe₂, confirming their enhanced sensing abilities. Charge density difference analysis shows significant charge redistribution, with CrTe₂ experiencing stronger charge transfer effects than CrSe₂. Variations in electrostatic potential and work function further demonstrate the higher sensitivity of CrTe₂, particularly in its defective and doped forms, confirming its status as a superior material for gas sensing applications.

**Keywords:** Monolayers, DFT, Adsorption, Sensing, Work function

## Introduction

The rapid growth of industrialization and urbanization is causing a rise in common pollutants and industrial waste gases, resulting in health hazards and environmental problems. This causes an increased level of certain pollutants, such as carbon monoxide (CO), nitrogen oxide (NO), sulfur dioxide ($SO_2$), and methane ($CH_4$) [1–4]. Among the mentioned pollutants, $SO_2$ is a colorless, irritating gas with a pungent odor that is harmful to the respiratory system and human health[5]. The affinity of CO for hemoglobin is stronger than that of oxygen; therefore, when CO is inhaled, it prevents oxygen from reaching various organs in the body, leading to severe damage to tissues and organs[6]. NO is the most prominent toxic gas, primarily produced by chemical and biological processes, posing significant risks to human and animal health. For instance, the Occupational Safety and Health Administration (OSHA) sets exposure limits of 25 ppm for NO over an 8-hour work shift and 50 ppm for $NH_3$. Exposure beyond the mentioned limits can cause severe breathing difficulties, irritation of the skin, eyes, nose, and throat, and may even lead to death[7]. Large emissions of $CH_4$, a potent greenhouse gas, are directly associated with environmental challenges, such as climate change[8].

These urgent issues have driven increased research into discovering more sustainable and efficient sensors for detecting trace amounts of CO, NO, $SO_2$, and $CH_4$ gases. Researchers are prioritizing advancements in detection technologies to improve sensitivity and accuracy. The focus is on developing materials that are both effective and environmentally friendly. Researchers are also exploring novel approaches to detect these gases at lower concentrations. This growing interest aims to address the environmental and health risks posed by these pollutants. The scientific community has been dedicated to discovering new materials for gas sensing applications, including metal oxide semiconductors and graphene [9,10].

Due to the zero-band gap of graphene[11] and the high operating temperatures of metal oxides[12], researchers have begun exploring other two-dimensional (2D) materials for sensing applications.

Transition metal dichalcogenides (TMDs) have gained significant attention for their remarkable physical and chemical properties[13,14] They offer benefits such as low energy consumption, fast recovery times, and high selectivity and sensitivity, making them ideal for gas sensing. With their semiconductor and semi-metal characteristics, TMDs are widely utilized in optoelectronics and spintronics. Their tunable properties and high surface area further enhance their potential in advanced sensing and electronic applications[15,16]. Several techniques exist to improve the sensitivity and selectivity of sensing materials, including doping, surface functionalization, vacancy creation, and strain engineering[17]. Spin-polarized DFT calculations have been used to study the sensing of $NH_3$, $NO$, and $NO_2$ using both pristine and substituted $MoSe_2$ and $MoTe_2$. The results indicate that substituted $MoSe_2$ and $MoTe_2$ are promising materials for detecting toxic gases[18]. Wu and his colleagues studied the effect of decorating $MoS_2$ monolayers with Pd clusters, finding that it enhances the detection of toxic gases like $NO_2$ and $SO_2$ compared to $MoS_2$ decorated with Cu, Au, and Ta, as demonstrated through DFT calculations [19]. Recently, many other studies explored the potential of different TMDs in gas-sensing applications [20–26].

Among various TMDs, chromium diselenide ($CrSe_2$) [27–29] and chromium ditelluride ($CrTe_2$) [30,31] are highly promising 2D materials with significant potential for applications in electronic devices, optoelectronic devices, catalysis, and sensors. Recently, Zhu et. Al [32] investigated sensing of $CH_4$, $H_2S$, and $CO$ with pristine and transition metal-substituted $CrSe_2$ using DFT calculations, and the results show that, except for $CH_4$, other gases are chemisorbed on the substituted $CrSe_2$.

To the best of our knowledge, there is no available literature on the sensing of toxic gases using pristine and substituted $CrTe_2$ monolayers. To fill this gap, this study uses DFT calculations to assess the gas-sensing properties of $CrSe_2$ and $CrTe_2$ monolayers. The vacancy-induced $CrTe_2$ ($V_{Te}$-$CrTe_2$) shows the strongest adsorption for CO, NO, and $SO_2$, while $CrSe_2$ demonstrates moderate, reversible binding. Ge, Sb, and Sn doping in $CrSe_2$ and CrTe further enhances sensing performance. Electronic structure, charge density, and electrostatic potential analyses highlight significant changes, especially in doped and imperfect $CrTe_2$, confirming its superior sensitivity and potential as an effective gas sensor.

## Computational details

All calculations in this study were performed using the Vienna Ab initio Simulation Package (VASP), based on density functional theory (DFT)[33,34]. The exchange and correlation functions were described using the generalized gradient approximation (GGA) with the Perdew–Burke–Ernzerhof (PBE) functional[35,36]. The Projector Augmented-Wave (PAW) pseudopotentials were employed to model the interaction between the electrons and ions[37,38]. Van der Waals interactions were accounted for, with the D3-Grimme correction (DFT-D3) applied to enhance the accuracy of long-range dispersion forces between the solid adsorbents and gas molecules [39,40]. We used a 4 x 4 x 1 supercell of pristine, defective, and elemental-substituted $CrSe_2$ and $CrTe_2$ monolayers. A vacuum space of 20 Å was introduced to prevent interlayer interactions in the Z direction. The Monkhorst-Pack (MP) method was used for k-point sampling to simulate Brillouin zone integration, with a 3×3×1 grid for geometric optimization and a 5×5×1 grid for electronic analysis. The charge transfer between the adsorbed gas molecules and the $CrSe_2$ or $CrTe_2$ monolayers was calculated using Bader charge analysis[41]. The adsorption energy of gas molecules on the monolayer sheets was determined using the following formula:

$$E_{ads} = E_{monolayer+gas} - E_{monolayer} - E_{gas}$$

Where the first, second, and third terms represent the total energy of the $CrX_2$ (X=Se, Te) monolayers after gas adsorption, bare monolayers, and the energy of the gas molecule, respectively.

## Results and discussion

Figure 1 illustrates the optimized structures of $CrSe_2$, $CrTe_2$, $V_{Se}$-$CrSe_2$, and $V_{Te}$-$CrTe_2$ monolayers. After the optimization of the $CrSe_2$ ($V_{Se}$-$CrSe_2$), the Cr–Se bond lengths are found to be 2.52 (2.56) Å, while the Cr–Se–Cr bond angles are 53.5º (53.2º). Similarly, in $CrTe_2$ ($V_{Te}$-$CrTe_2$), the Cr–Te bond lengths were 2.68 (2.71) Å and the corresponding Cr–Te–Cr bond angles of 53.6º (54.8º). The presence of Se/Te vacancies in $V_{Se}$-$CrSe_2$ and $V_{Te}$-$CrTe_2$ causes slight variations in bond lengths and angles. The calculated geometrical parameters for $CrSe_2$ and $CrTe_2$ are in good agreement with previously reported results [42,43].

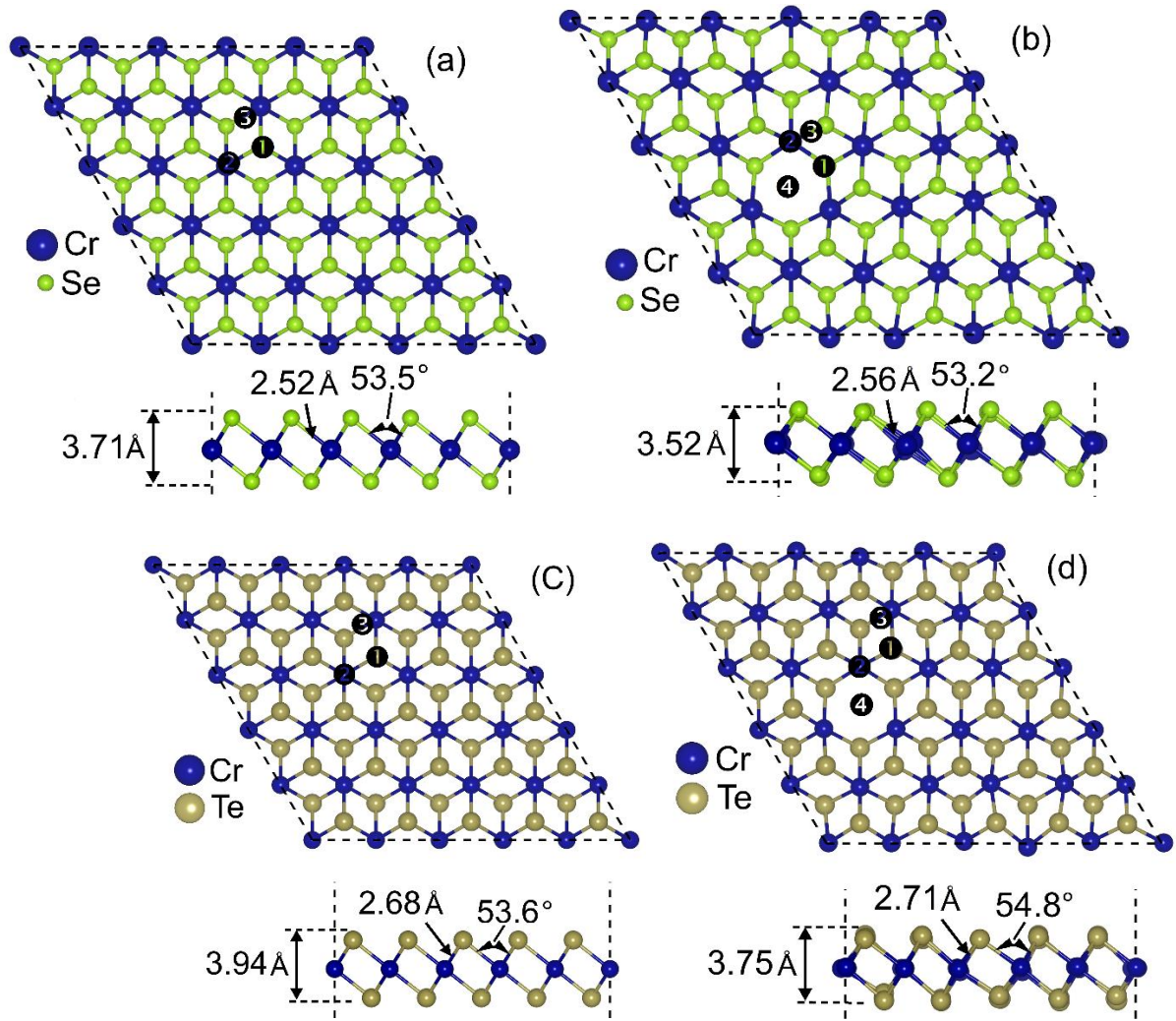

**Figure 1:** Optimized structures of (a) CrSe₂, (b) V_Se-CrSe₂, (c) CrTe₂, (d) V_Te-CrTe₂ monolayers.

The electronic properties are studied through partial density of states (PDOS). The PDOS plots in Figure 2 reveal a finite density of states at the Fermi level for CrSe₂, V_Se-CrSe₂, CrTe₂, and V_Te-CrTe₂ monolayers, confirming their metallic nature. In all the systems, the Cr (d) orbitals dominate the electronic states near the Fermi level, while Se/Te (p) states contribute less significantly. The introduction of V_Se/V_Te vacancies alters the PDOS but does not change the bandgap, preserving metallicity while introducing defect-induced states that may impact conductivity and electron scattering. These changes suggest that V_Se/V_Te vacancies impact the electronic properties of V_Se-CrSe₂ and V_Te-CrTe₂ monolayers.

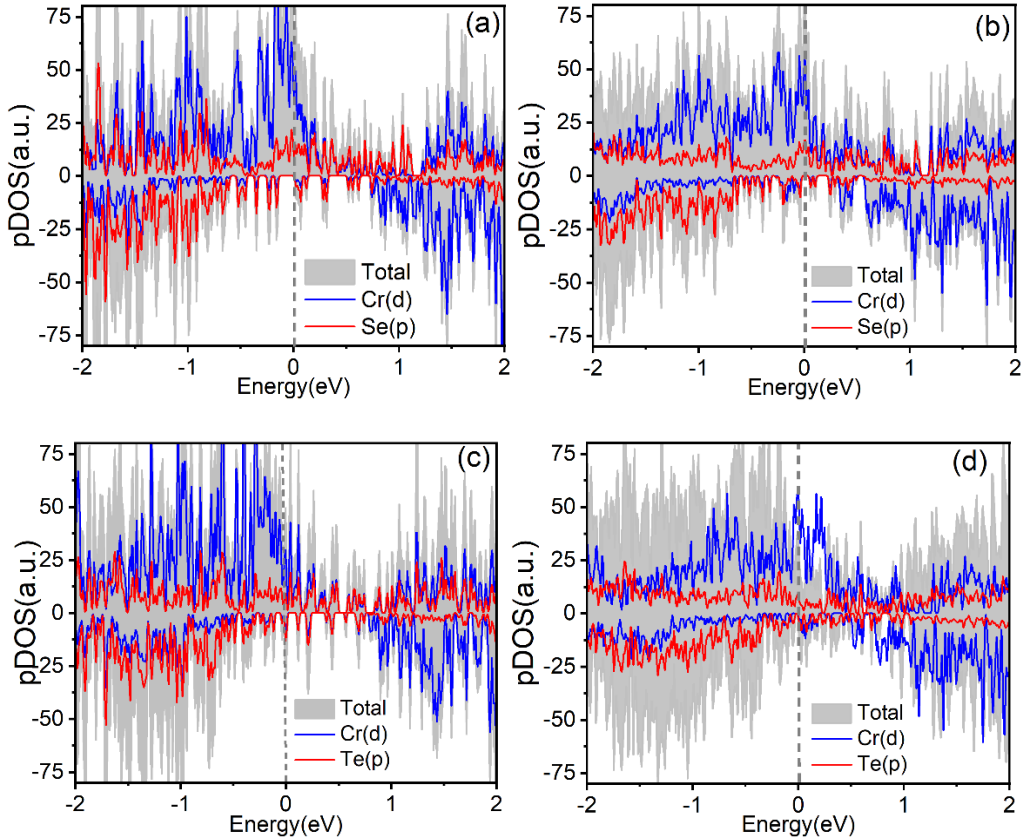

**Figure 2**: Projected density of states (PDOS) plots for (a) CrSe₂, (b) V_Se-CrSe₂, (c) CrTe₂, and (d) V_Te-CrTe₂ monolayers.

**Adsorption of gases on CrSe₂ and CrTe₂ monolayers:**

To determine the most stable structures of the gas adsorbed CrSe₂ and CrTe₂, we have considered all the possible adsorption sites. Additionally, various possible orientations of the gas molecules were examined. Table 1 presents key adsorption parameters, including the binding distance ($\Delta d_{S-m}$), $E_{ads}$ and net charge transfer ($\Delta\rho$), which helps evaluate the strength and nature of the interactions between the gas molecules and the CrSe₂ and CrTe₂ surfaces. Figure 3 presents the optimized structures of CO, NO, SO₂, and CH₄ adsorbed on CrSe₂ and CrTe₂ monolayers.

| System | CO | | | NO | | | SO$_2$ | | | CH$_4$ | | |
|---|---|---|---|---|---|---|---|---|---|---|---|---|
| | $\Delta d_{S\text{-}m}$ (Å) | $E_{ads}$ (eV) | $\Delta\rho$ (e) | $\Delta d_{S\text{-}m}$ (Å) | $E_{ads}$ (eV) | $\Delta\rho$ (e) | $\Delta d_{S\text{-}m}$ (Å) | $E_{ads}$ (eV) | $\Delta\rho$ (e) | $\Delta d_{S\text{-}m}$ (Å) | $E_{ads}$ (eV) | $\Delta\rho$ (e) |
| CrSe$_2$ | 3.48 | -1.13 | -0.65 | 1.86 | -1.17 | - | 3.57 | -0.90 | - | 3.27 | -1.12 | -0.67 |
| V$_{Se}$-CrSe$_2$ | 2.78 | -0.82 | -0.55 | 1.71 | -1.22 | -0.87 | 2.01 | -0.61 | -0.57 | 2.11 | -0.13 | - |
| Ge-CrSe$_2$ | 2.81 | -0.17 | -0.56 | 2.45 | -0.33 | - | 3.12 | -0.29 | - | 3.25 | -0.33 | - |
| CrTe$_2$ | 2.48 | -0.23 | -0.49 | 2.64 | -1.05 | - | 2.47 | -0.45 | - | 2.89 | -0.26 | - |
| V$_{Te}$-CrTe$_2$ | 2.29 | -1.52 | -0.74 | 2.14 | -1.79 | - | 2.98 | -1.61 | - | 2.44 | -0.27 | - |
| Sb-CrTe$_2$ | 4.08 | -0.78 | -0.32 | 2.85 | -1.46 | - | 3.27 | -1.01 | -0.64 | 3.11 | -0.78 | -0.92 |
| Sn-CrTe$_2$ | 3.85 | -1.19 | - | 2.38 | -1.19 | -0.94 | 3.01 | -1.34 | - | 2.48 | -0.11 | - |

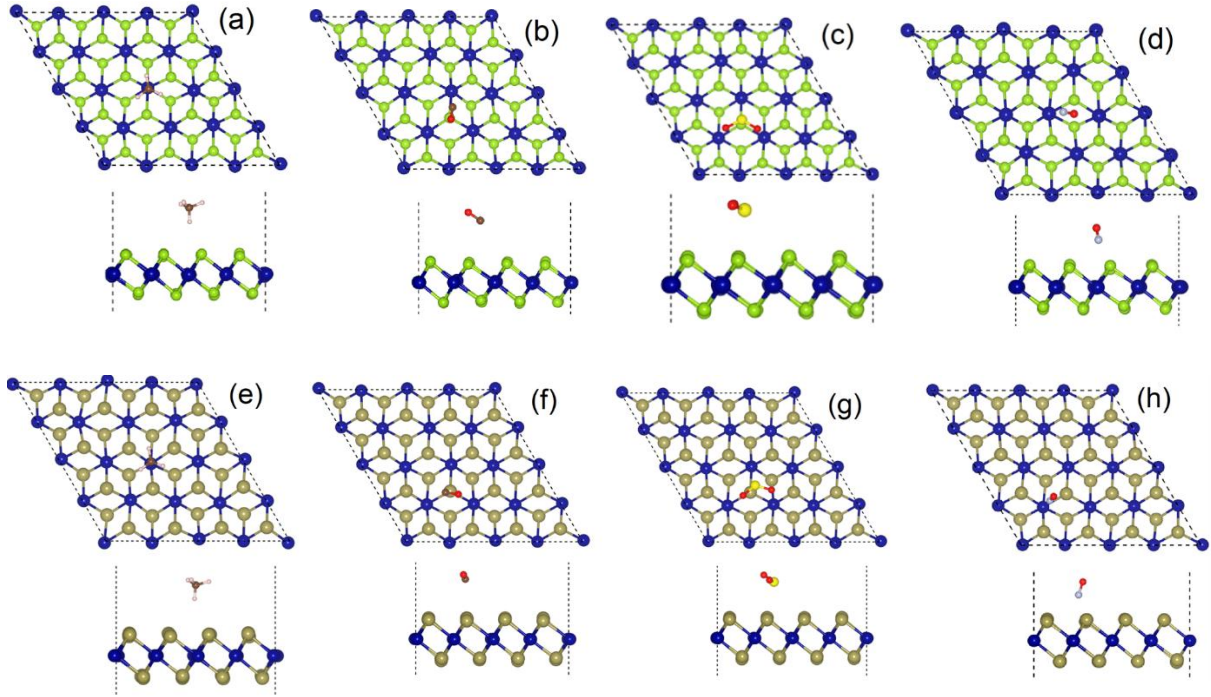

**Figure 3**: The most stable configurations of (a) CrSe$_2$-CH$_4$, (b) CrSe$_2$-CO, (c) CrSe$_2$-SO$_2$, (d) CrSe$_2$-NO, (e) CrTe$_2$-CH$_4$, (f) CrTe$_2$-CO, (g) CrTe$_2$-SO$_2$, (h) CrTe$_2$-NO.

The adsorption characteristics of CO, NO, SO$_2$, and CH$_4$ on CrSe$_2$ and CrTe$_2$ reveal distinct interaction trends. Upon CO adsorption, CrSe$_2$ exhibits a binding distance of 3.48 Å with an $E_{ads}$ value of -1.13 eV and a charge transfer of -0.65 e, indicating moderate interaction. In contrast,

CrTe₂ shows a shorter binding distance of 2.48 Å and significantly weaker $E_{ads}$ of -0.23 eV, suggesting a weaker affinity for CO. The adsorption of NO on CrSe₂ results in a binding distance of 1.86 Å and an $E_{ads}$ of -1.17 eV, whereas CrTe₂ exhibits a larger binding distance of 2.64 Å but retains a relatively strong $E_{ads}$ of -1.05 eV. The SO₂ adsorption on CrSe₂ occurs at 3.57 Å with an $E_{ads}$ of -0.90 eV, whereas CrTe₂ has a slightly shorter binding distance of 2.47 Å and a lower $E_{ads}$ of -0.45 eV, indicating a weaker interaction compared to CrSe₂. For CH₄, both CrSe₂ and CrTe₂ exhibit weak adsorption, with binding distances of 3.27 and 2.89 Å, respectively, and $E_{ads}$ of -1.12 and -0.26 eV. The results indicate that CrSe₂ generally exhibits stronger interactions with gas molecules than CrTe₂, particularly for CO and NO, suggesting that it may be more effective for gas adsorption applications.

The adsorption of gas molecules CO, CH₄, and SO₂ on pristine CrSe₂ and NO on CrTe₂ monolayer significantly influences their electronic structure, as shown in Figure 4.

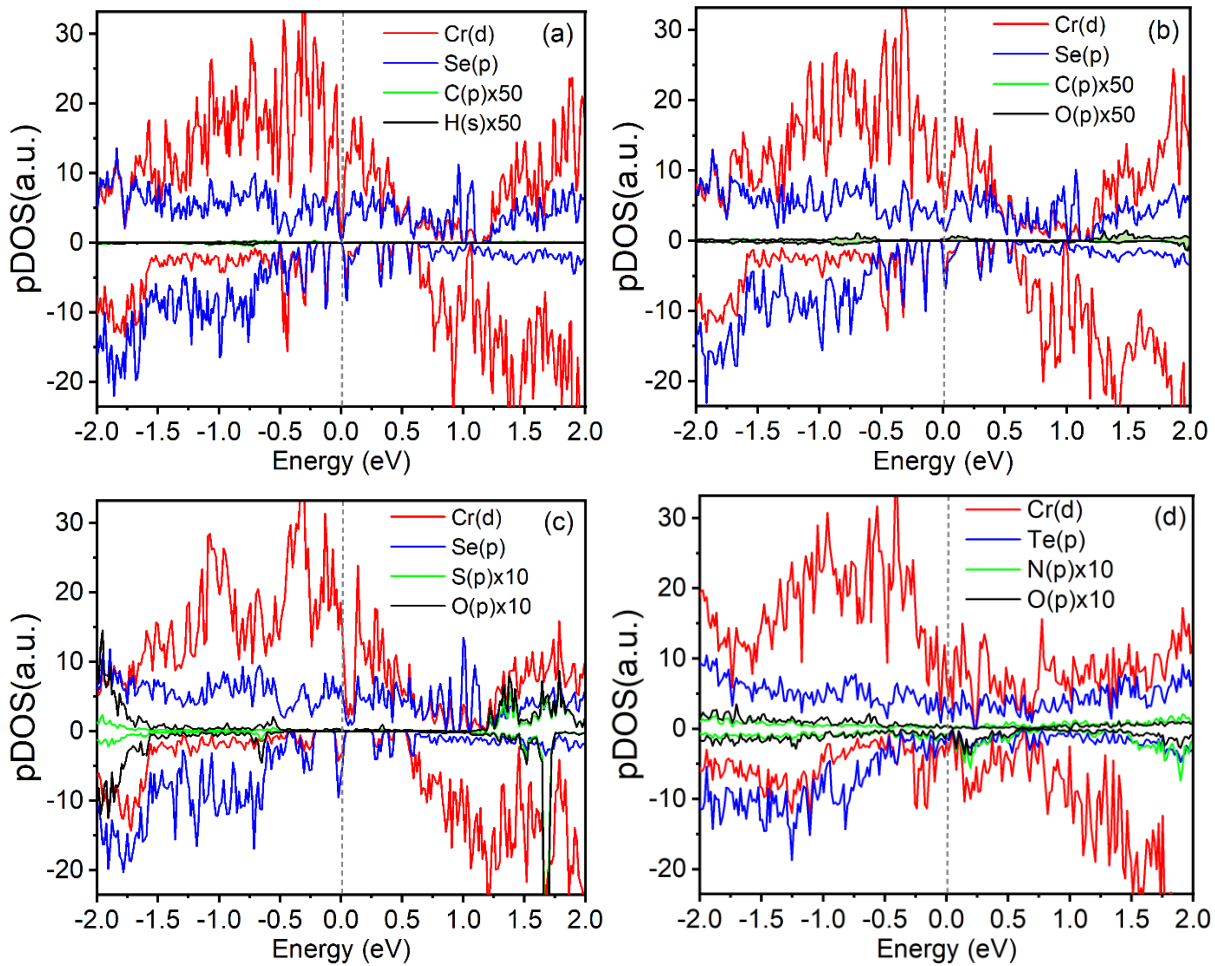

**Figure 4:** Electronic structure of pristine $CrSe_2$ and $CrTe_2$ upon the adsorption of incident gases; (a) $CrSe_2$-$CH_4$, (b) $CrSe_2$-CO, (c) $CrSe_2$-$SO_2$, and (d) $CrTe_2$-NO.

In these systems, the Cr (d) orbitals remain dominant near the Fermi level, but the interaction with the incident gas molecules introduces new states. For example, in $CrSe_2$-CO and $CrTe_2$-NO, hybridization between the Cr d-orbitals and the gas molecule's 'p' orbitals leads to additional states near the Fermi level, indicating charge transfer and possible chemisorption. The effect of adsorption varies depending on the gas; $CH_4$ shows weak interaction with minimal DOS changes, while molecules like NO and $SO_2$ create stronger modifications by introducing localized states near the conduction or valence band edges.

**Adsorption of gas molecules on $V_{Se}$-$CrSe_2$ and $V_{Te}$-$CrTe_2$:**

The adsorption of the gas molecules on $V_{Se}$-$CrSe_2$ and $V_{Te}$-$CrTe_2$ monolayers is depicted in Figure 5. The introduction of Se (Te) vacancies significantly alters the adsorption behaviours of $V_{Se}$-$CrSe_2$ and $V_{Te}$-$CrTe_2$ towards the incident gas molecules. Upon CO adsorption, $V_{Se}$-$CrSe_2$ exhibits a shorter binding distance of 2.78 Å compared to $CrSe_2$, with a weaker $E_{ads}$ of -0.82 eV and a charge transfer of -0.55e, indicating moderate interaction. In contrast, $V_{Te}$-$CrTe_2$ demonstrates a stronger interaction with CO, characterized by a shorter binding distance of 2.29 Å and significantly stronger $E_{ads}$ of -1.52 eV, accompanied by a substantial charge transfer of -0.74 e. The NO adsorption on $V_{Se}$-$CrSe_2$ occurs at a binding distance of 1.71 Å with $E_{ads}$ of -1.22 eV and a charge transfer of -0.87 e, indicating a strong interaction.

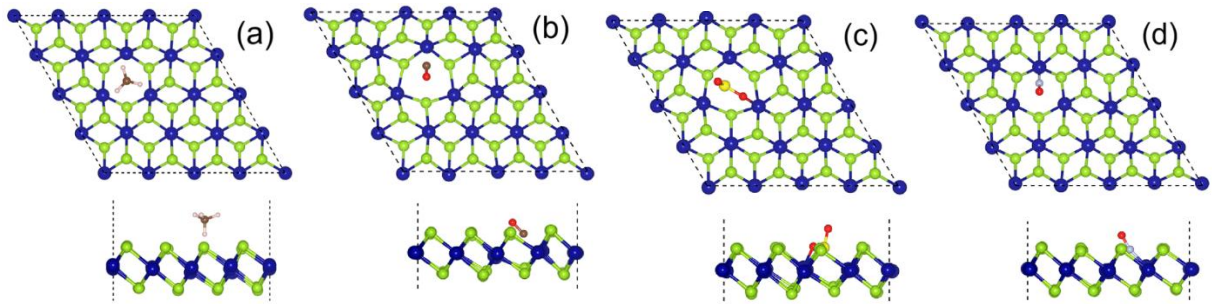

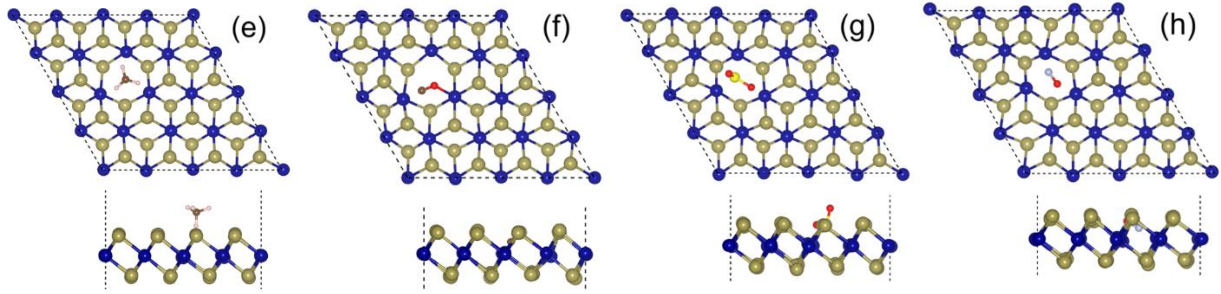

Figure 5: Adsorption of gas molecules on (a) $V_{Se}$-CrSe$_2$-CH$_4$, (b) $V_{Se}$-CrSe$_2$-CO, (c) $V_{Se}$-CrSe$_2$-SO$_2$, and (d) $V_{Se}$-CrSe$_2$-NO, (e) $V_{Te}$-CrTe$_2$-CH$_4$, (f) $V_{Te}$-CrTe$_2$-CO, (g) $V_{Te}$-CrTe$_2$-SO$_2$, and (h) $V_{Te}$-CrTe$_2$-NO.

Similarly, $V_{Te}$-CrTe$_2$ exhibits even stronger adsorption towards NO, with a shorter binding distance of 2.14 Å and the strongest $E_{ads}$ of -1.79 eV, signifying enhanced gas capture capability. For SO$_2$ adsorption, $V_{Te}$-CrTe$_2$ shows a binding distance of 2.01 Å and $E_{ads}$ of -0.61 eV, whereas $V_{Te}$-CrTe$_2$ presents a binding distance of 2.98 Å and significantly stronger $E_{ads}$ of -1.61 eV for SO$_2$ molecules. CH$_4$ adsorption remains weak on both vacancy-engineered surfaces, with $V_{Se}$-CrSe$_2$ showing a binding distance of 2.11 Å and $E_{ads}$ of -0.13 eV, while $V_{Te}$-CrTe$_2$ has a slightly shorter distance of 2.44 Å and $E_{ads}$ of -0.27 eV. The results indicate that the vacancy significantly enhances gas adsorption, especially for CO and NO, with $V_{Te}$-CrTe$_2$ exhibiting the strongest interactions among all configurations.

Both $V_{Se}$-CrSe$_2$ and $V_{Te}$-CrTe$_2$ enhance gas adsorption as compared to the pristine monolayers due to the presence of unsaturated metal sites. Improvement in the adsorption mechanism is further explored through PDOS plots. In $V_{Se}$-CrSe$_2$-CO, $V_{Se}$-CrSe$_2$-SO$_2$, and $V_{Te}$-CrTe$_2$-CO systems, the PDOS plots reveal new mid-gap states contributed by the p orbitals of the Se/Te vacancies. These vacancies create reactive sites that promote stronger interactions with the incident gas molecules, as shown in Fig. 6. For example, in $V_{Se}$-CrSe$_2$-NO and $V_{Te}$-CrTe$_2$-CO, the gas molecule's electronic states strongly overlap with the Cr d-states, shifting the Fermi level and potentially altering the material's conductivity. Such vacancy-induced modifications suggest enhanced gas sensitivity, making these materials promising for gas sensing or catalytic applications.

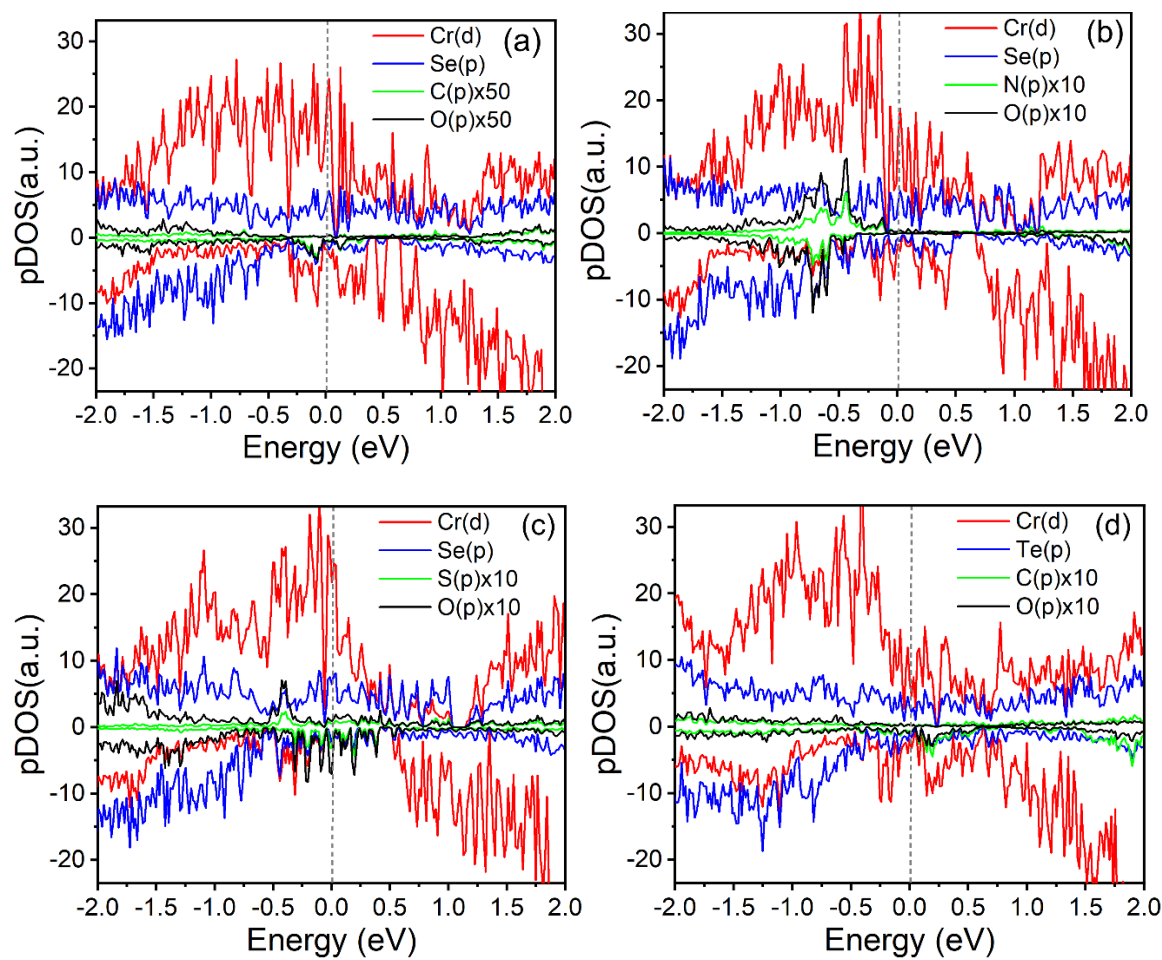

Figure 6. Influence of gas molecule adsorption on the electronic structure of (a) ($V_{Se}$-CrSe$_2$-CO, (b) $V_{Se}$-CrSe$_2$-NO, (c) $V_{Se}$-CrSe$_2$-SO$_2$, and (d) $V_{Te}$-CrTe$_2$-CO.

**Adsorption of gas molecules on substituted CrSe$_2$ and CrTe$_2$ monolayers**

We further explore the influence of the selected substituents on the sensing properties of CrSe$_2$ and CrTe$_2$. For this purpose, selected elements, such as germanium (Ge), antimony (Sb), and tin (Sn), are considered to improve the sensing mechanism. The optimized geometries of substituted systems are shown in Figure S1. These elements have been selected for their cost-effectiveness compared to precious metals such as platinum (Pt), gold (Au), and palladium (Pd). In addition to their affordability and availability, Ge, Sb, and Sn exhibit distinctive reactivity, making them highly suitable for diverse catalytic applications. We observe that the substitution of Ge in CrSe$_2$ (Ge-CrSe$_2$), and Sb and Sn in CrTe$_2$ (Sb-CrTe$_2$, Sn-CrTe$_2$) alter their adsorption behaviour toward CO, NO, SO$_2$, and CH$_4$ molecules. The optimized geometries of substituted systems are shown in Figure 7.

In the case of Ge-CrSe$_2$, CO adsorption occurs at a binding distance of 2.81 Å with weak $E_{ads}$ of -0.17 eV and a charge transfer of 0.56 e, suggesting electronic interaction but weak physisorption. NO adsorption on Ge-CrSe$_2$ shows a binding distance of 2.45 Å with $E_{ads}$ of -0.33 eV, indicating a relatively weak interaction. Similarly, SO$_2$ adsorption results in a binding distance of 3.12 Å and $E_{ads}$ of -0.29 eV, while CH$_4$ adsorption occurs at 3.25 Å with $E_{ads}$ of -0.33 eV, both indicating weak physisorption.

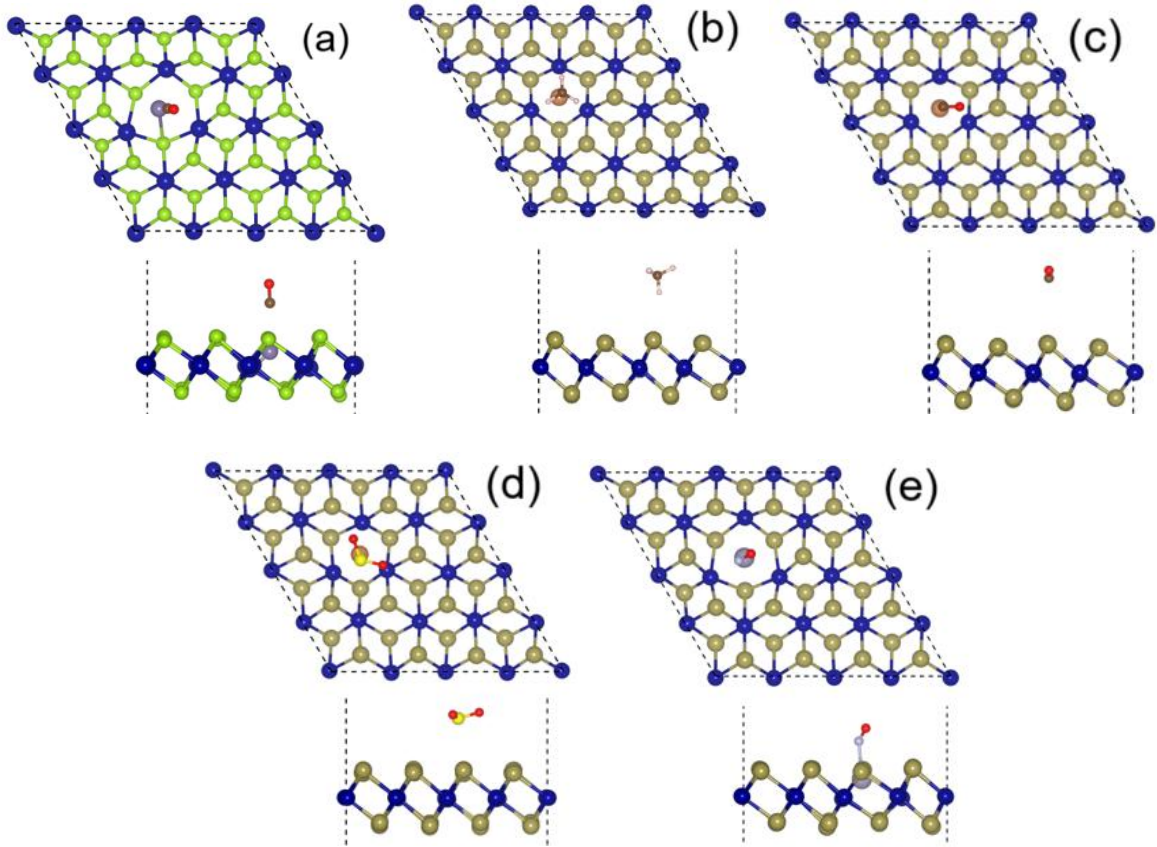

**Figure 7**: Adsorption of gas molecules on (a) Ge-CrSe$_2$-CO, (b) Sb-CrTe$_2$-CH$_4$, (c) Sb-CrTe$_2$-CO, (d) Sb-CrTe$_2$-SO$_2$, (d) Sn-CrTe$_2$-NO

For Sb-CrTe$_2$, CO adsorption follows at a notably larger binding distance of 4.08 Å with $E_{ads}$ of -0.78 eV and a charge transfer of 0.32 e, suggesting appropriate interaction with significant electron exchange. NO adsorption on Sb-CrTe$_2$ is stronger, with a binding distance of 2.85 Å and $E_{ads}$ of -1.46 eV, implying a strong interaction. SO$_2$ binds at 3.27 Å with an adsorption energy of -1.01 eV and a charge transfer of 0.64 e, while CH$_4$ binds at 3.11 Å with $E_{ads}$ of -0.78 eV and a charge transfer of -0.92 e, showing moderate interaction. Sn-CrTe$_2$ exhibits slightly stronger CO adsorption than Ge-CrSe$_2$, with a binding distance of 3.85 Å and $E_{ads}$ of -1.19 eV. NO adsorption occurs at 2.38 Å with an adsorption energy of -1.19 eV and a charge transfer of 0.94 e, indicating strong binding. $SO_2$ adsorption shows a binding distance of 3.01 Å with $E_{ads}$ of -1.34 eV, while $CH_4$ binds at 2.48 Å with weak $E_{ads}$ of -0.11 eV. These results indicate that substitution affects adsorption properties differently, with Sb-CrTe$_2$ and Sn-CrTe$_2$ showing relatively strong interactions for NO and $SO_2$, while Ge-CrSe$_2$ exhibits weaker adsorption across all gas molecules.

Elemental substitution further modifies the electronic structure of gas adsorbed Ge-CrSe$_2$, Sb-CrTe$_2$, and Sn-CrTe$_2$ monolayers. In the systems, Ge-CrSe$_2$-CO, Sb-CrTe$_2$-CH$_4$, and Sn-CrTe$_2$-NO, the PDOS exhibits altered hybridization patterns due to the introduction of new electronic states. For instance, Sb-CrTe$_2$ increases p-orbital contributions, leading to different charge transfer behaviour with adsorbed gas molecules. Similarly, in Ge-CrSe$_2$, the Ge p-orbitals influence gas interactions, particularly with CO, leading to modified energy levels near the Fermi level. These variations highlight that substitution can be a useful strategy for tailoring gas adsorption properties in Cr-based materials.

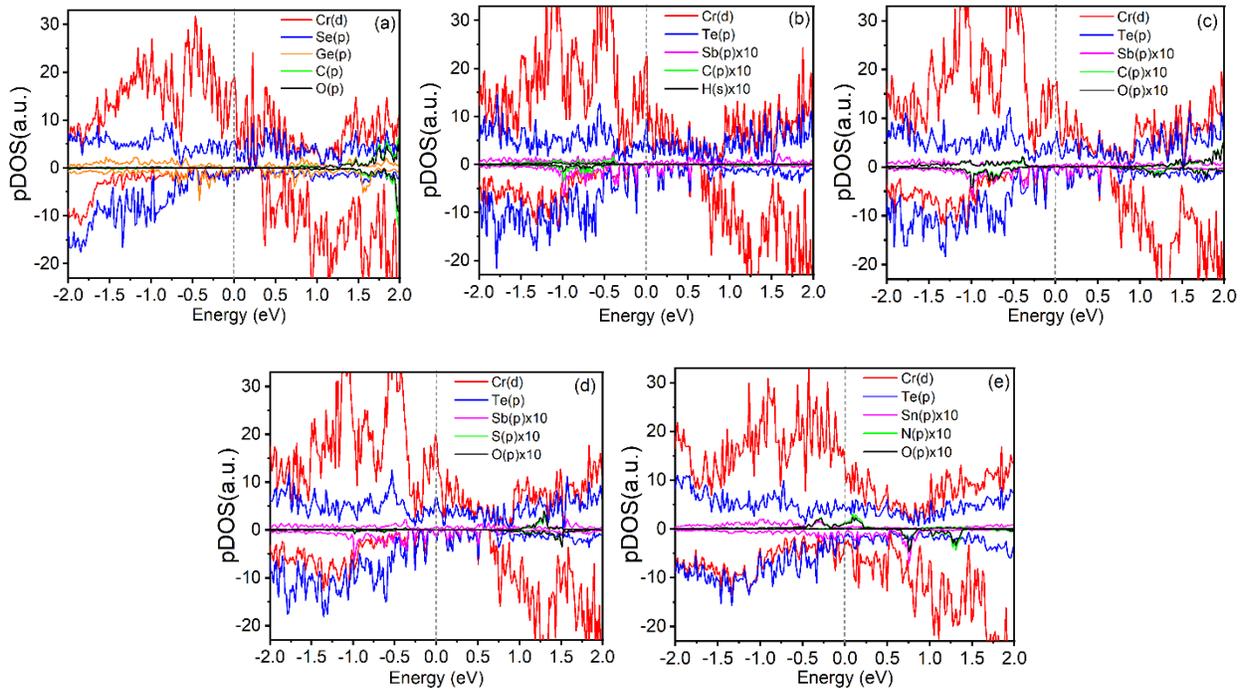

Figure 8: PDOS plots of (a) Ge-CrSe$_2$-CO, (b) Sb-CrTe$_2$-CH$_4$, (c) Sb-CrTe$_2$-CO, (d) Sb-CrTe$_2$-SO$_2$, (d) Sn-CrTe$_2$-NO.

**Charge density difference analysis**

The charge density difference plots for pristine, vacancy-induced and substituted systems are given in Figure 9. $CrSe_2$ and $CrTe_2$ adsorbed with $CH_4$, $CO$, and $SO_2$, reveal moderate charge redistribution at the adsorption sites (Figure 9). The yellow and cyan regions indicate electron accumulation and depletion, respectively. In $CrSe_2$-CO and $CrTe_2$-NO, strong charge redistribution occurs near the gas molecule and the Cr sites, suggesting notable charge transfer. In contrast, weak physisorption is observed in $CrSe_2$-$CH_4$, where minimal charge accumulation occurs around the gas molecule. These results indicate that CO and NO interact more strongly with $CrSe_2$ and $CrTe_2$, leading to potential electronic structure modifications. For $V_{Se}$-$CrSe_2$-CO, $V_{Se}$-$CrSe_2$-$SO_2$, and $V_{Te}$-$CrTe_2$-CO, the charge density plots show enhanced electron redistribution due to the presence of vacancies. The vacancy sites act as active centres, leading to significant electron accumulation around the gas molecules, particularly in $V_{Se}$-$CrSe_2$-$SO_2$, and $V_{Te}$-$CrTe_2$-CO. This suggests strong adsorption behaviour, with the gas molecules donating or withdrawing charge from the defective surface. The localized charge transfer is more pronounced compared to pristine structures, demonstrating that vacancies enhance the material's reactivity towards gas adsorption.

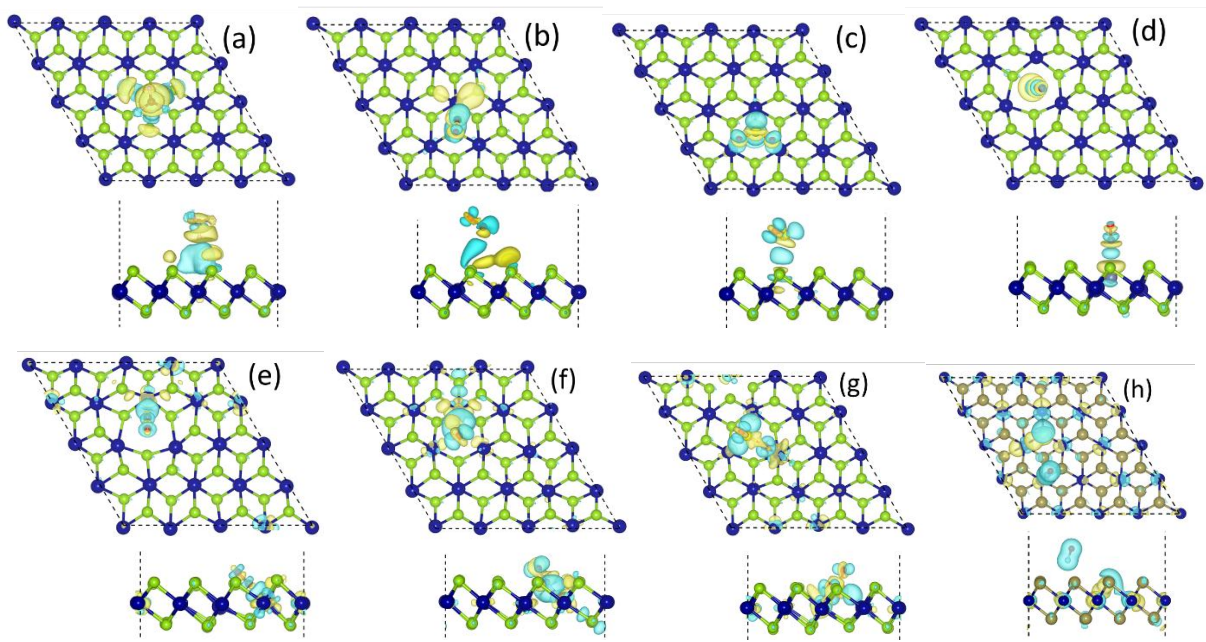

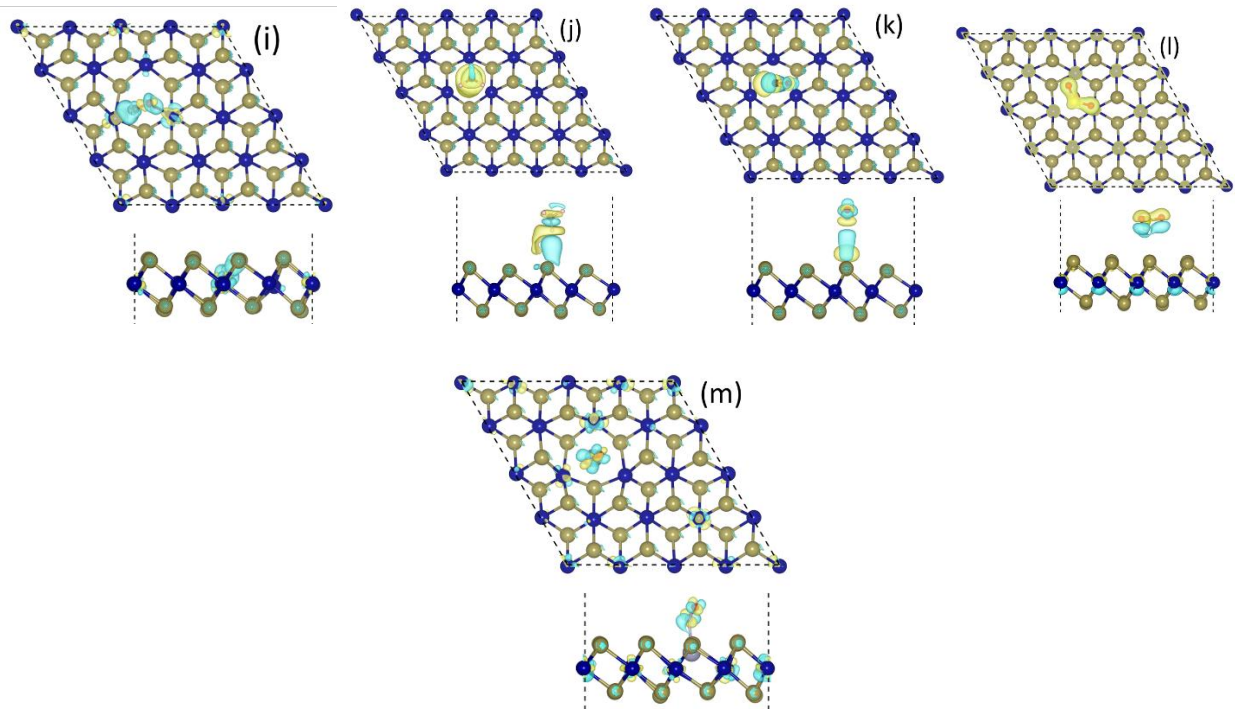

Figure 9: Charge density difference plot of gas molecules on pristine, vacancy-induced, and substituted systems. (a) $CrSe_2$-$CH_4$, (b) $CrSe_2$-CO, (c) $CrSe_2$-$SO_2$, and (d) Ge-$CrSe_2$-CO, (e) $V_{Se}$-$CrSe_2$-CO, (f) $V_{Se}$-$CrSe_2$-NO, (g) $V_{Se}$-$CrSe_2$-$SO_2$, (h) $CrTe_2$-NO, (i) $V_{Te}$-$CrTe_2$-CO, (j) Sb-$CrTe_2$-$CH_4$, (k) Sb-$CrTe_2$-CO, (l) Sb-$CrTe_2$-$SO_2$, (m) Sn-$CrTe_2$-NO.

In substituted systems, such as Ge-$CrSe_2$-CO, Sb-$CrTe_2$-CO, and Sn-$CrTe_2$-NO, the charge density difference plots reveal distinct electronic effects due to doping. For instance, in Ge-$CrSe_2$-CO and Sb-$CrTe_2$-CO, charge redistribution is more delocalized, suggesting altered bonding characteristics between the gas molecules and the surfaces. Sn-$CrTe_2$-NO exhibits noticeable electron accumulation around the NO molecule, indicating significant charge transfer, which may influence the adsorption energy and sensing properties. These modifications suggest that substitution tuning can optimize Cr-based materials for specific gas interactions.

**Electrostatic potential**

The electrostatic potential profile for pristine, vacancy-induced, and substituted monolayers adsorbed with gas molecules (Figure 10) demonstrates a notable variation in potential across the interface. The deep potential well within the $CrSe_2$ indicates strong internal electrostatic interactions, while the potential near the vacuum level shows shifts depending on the adsorbed gas molecule. Compared to $CrSe_2$, $V_{Se}$-$CrSe_2$-CO, $V_{Se}$-$CrSe_2$-NO, and $V_{Se}$-$CrSe_2$-$SO_2$ exhibit a reduction in the potential barrier, suggesting enhanced charge transfer due to the presence of vacancies. Ge-CrSe₂ slightly modifies the potential drop, indicating a controlled tuning effect. These shifts in potential imply that defects and substitution enhance gas adsorption by facilitating charge redistribution, making imperfect and substituted CrSe₂ promising for gas-sensing applications.

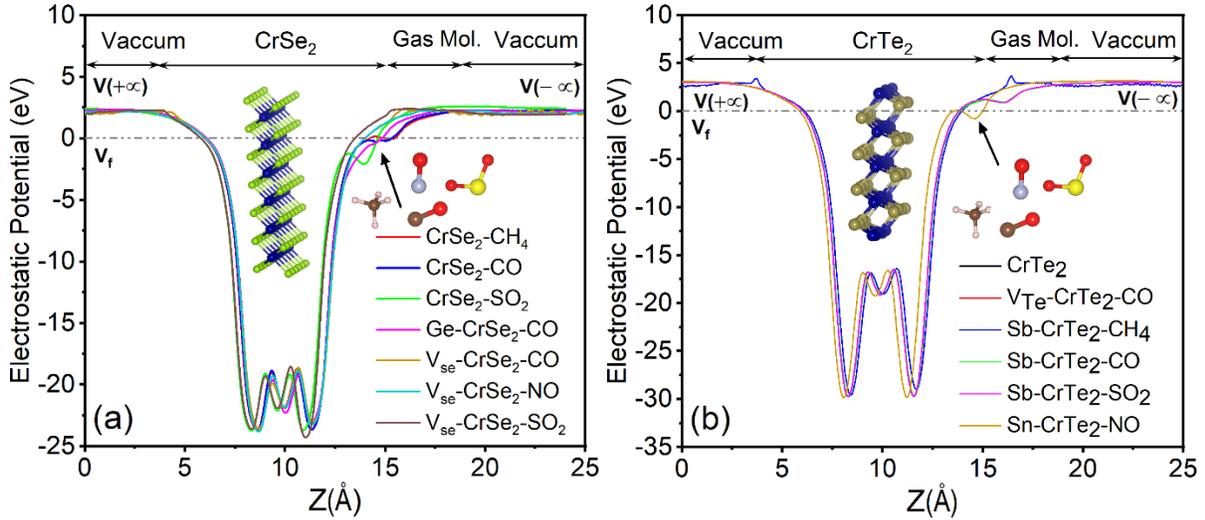

Figure 10: Electrostatic potential with and without adsorption of gas molecules on pristine, defective, and substituted CrSe₂ and CrTe₂ monolayers.

In the case of CrTe₂ monolayers (Figure 10b), the electrostatic potential profile follows a similar trend but with deeper potential wells, suggesting stronger internal electrostatic interactions. The pristine CrTe₂ shows a symmetric potential drop, whereas $V_{Te}$-CrTe₂-CO and $V_{Te}$-CrTe₂-SO₂ demonstrate an asymmetric shift, indicating enhanced gas interaction due to the presence of vacancies. Substituted systems, such as Sb-CrTe₂-CO and Sn-CrTe₂-NO, show moderate variations in the vacuum level potential, which can be attributed to altered charge transfer properties. The changes in electrostatic potential upon gas adsorption suggest that defect engineering and elemental substitution effectively modulate the electronic properties of CrTe₂, enhancing its sensitivity to gas molecules and making it a viable candidate for gas-sensing applications. Overall, while both CrSe₂ and CrTe₂-based systems benefit from defect engineering and substitution, CrTe₂ exhibits a stronger response to gas adsorption due to deeper electrostatic potential wells and greater asymmetry in imperfect structures. This suggests that CrTe₂ may be more sensitive to gas molecules, whereas CrSe₂ could offer more stability and selectivity in sensing applications.

**Work function**

The work function ($\phi$) of CrSe$_2$ (marked with a red dashed line) is around 5.0 eV (Figure 11). Upon gas adsorption, variations in $\phi$ are observed, indicating charge transfer between the gas molecules and the CrSe$_2$ surface. Among the adsorbed gases, SO$_2$ exhibits the highest $\phi$ increase, suggesting strong electron withdrawal from CrSe$_2$. V$_{Se}$-CrSe$_2$ shows a slightly higher baseline $\phi$ (black dashed line), and gas adsorption further modifies its electronic structure. Ge-CrSe$_2$ leads to a notable increase in $\phi$, indicating enhanced charge redistribution upon gas interaction. Overall, the increase or decrease in $\phi$ depends on whether the adsorbed gas donates or withdraws electrons from the monolayer.

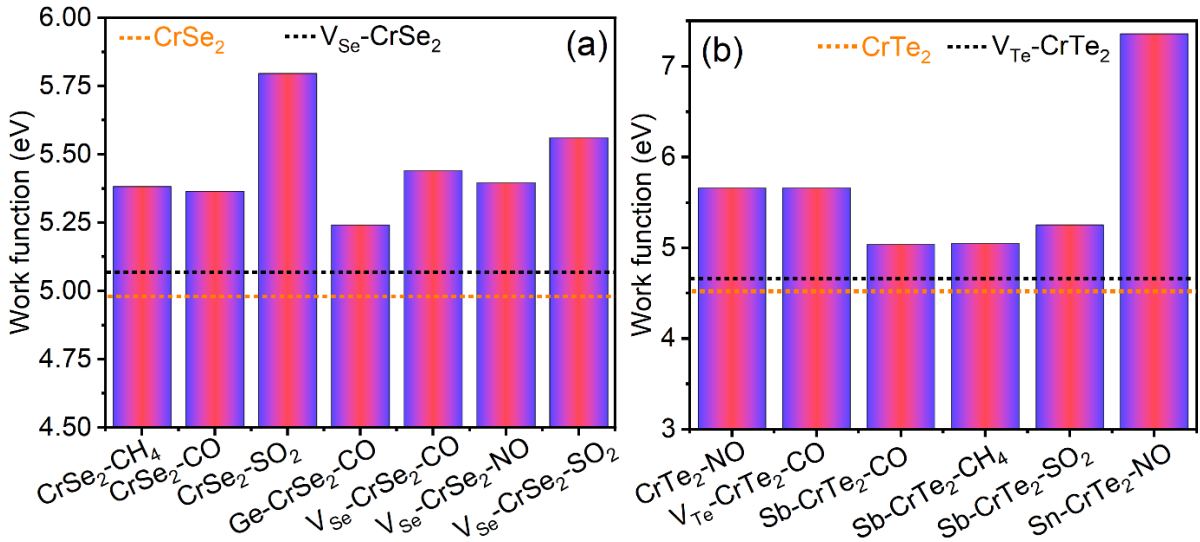

Figure 11: The work function of with and without adsorption of gas molecules on pristine, defective, and substituted CrSe$_2$ and CrTe$_2$ monolayers.

The $\phi$ value of CrTe$_2$ (red dashed line) is slightly lower than CrSe$_2$, indicating a different charge transfer capability. Gas adsorption leads to varying shifts in the $\phi$, with NO adsorption showing the most significant increase, suggesting strong electron affinity. V$_{Te}$-CrTe$_2$ exhibits a moderate increase in $\phi$ compared to its pristine form. Sb-CrTe$_2$ and Sn-CrTe$_2$ show significant variations in $\phi$, particularly upon NO and SO$_2$ adsorption, indicating enhanced sensitivity to these gas species. The larger shifts in $\phi$ for CrTe$_2$ suggest that it may be more responsive to gas molecules than CrSe$_2$. Both CrSe$_2$ and CrTe$_2$ exhibit $\phi$ modulation upon gas adsorption, with vacancy-induced and substituted systems showing enhanced sensitivity. However, CrTe$_2$ displays larger shifts in $\phi$, particularly for NO adsorption, making it a more promising candidate for gas sensing applications where $\phi$ variations serve as a detection mechanism. CrSe$_2$, on the other hand, provides more stability and selective interaction with specific gases, particularly SO$_2$. The choice between CrSe$_2$ and CrTe$_2$ would depend on the desired sensitivity and selectivity in gas-sensing applications.

Compared to previous studies, the current work on CrTe$_2$ and CrSe$_2$ monolayers, especially in their Se/Te vacancies and substituted forms, shows significantly enhanced gas adsorption capabilities similar to covalent triazine framework-1 (CTF-1) [44]. For instance, CrTe$_2$ exhibits notable chemisorption with adsorption energies of -1.52 eV for CO, -1.79 eV for NO, and -1.61 eV for SO$_2$, whereas the CTF-1 framework displays much weaker physisorption, with energies such as -0.32 eV for CO and -0.47 eV for SO$_2$. This contrast suggests that CrTe$_2$-based materials are more promising for highly sensitive electronic gas sensing, while the CTF-1 framework is better suited for reversible gas storage applications. Compared to Zn clusters reported in the literature[45], which show very strong SO$_2$ adsorption (up to -4.99 eV) but weak CO adsorption (-0.08 eV), the CrTe$_2$ and CrSe$_2$ monolayers studied here provide more balanced adsorption energies across multiple gases. Notably, CrTe$_2$ with vacancies or Sb/Sn doping maintains strong and consistent adsorption energies (-1.52 eV for CO, -1.79 eV for NO, and -1.61 eV for SO$_2$), making it an excellent candidate for selective and reusable gas sensing[46]. Furthermore, doped MoTe$_2$ reported adsorption energies up to -1.15 eV for SO$_2$ on Ag-doped MoTe$_2$, with weaker binding for CO and NO$_2$ (-0.32 to -0.84 eV). In comparison, the CrTe$_2$ monolayers in this work exhibit stronger and more uniform adsorption, reaching -1.79 eV for NO and -1.61 eV for SO$_2$, highlighting their superior sensitivity and potential for advanced gas sensing applications.

## Conclusion

Nanosensors play a critical role in environmental monitoring, industrial safety, and public health by enabling the detection of hazardous gases such as CO, NO, SO$_2$, and CH$_4$ at trace levels. This study presents a comprehensive density functional theory (DFT) investigation of the gas sensing capabilities of CrSe$_2$ and CrTe$_2$ monolayers, including their pristine, vacancy-induced, and substituted forms. Adsorption energy ($E_{ads}$) calculations reveal that Te-vacancy-induced CrTe$_2$ (V$_{Te}$-CrTe$_2$) exhibits the strongest gas binding, particularly for NO and SO$_2$, with values as low as -1.79 eV, while CrSe$_2$ shows moderate yet reversible binding suitable for sensing applications. Electronic structure analyses, including density of states (DOS), charge density difference, electrostatic potential, and work function ($\phi$), provide further insights into the sensing mechanisms.

Imperfect and Sb/Sn-doped $CrTe_2$ monolayers display substantial changes in DOS near the Fermi level and pronounced charge redistribution, indicating higher chemical reactivity and improved sensing response. Greater shifts in electrostatic potential and more significant variations in work function upon gas adsorption further reinforce the superior sensitivity of $CrTe_2$ compared to $CrSe_2$. Overall, while both $CrSe_2$ and $CrTe_2$ monolayers demonstrate potential as 2D gas-sensing materials, $CrTe_2$, especially in its vacancy-induced and doped configurations, emerges as a more effective and reliable candidate for detecting toxic gases due to its stronger binding affinities, enhanced electronic responsiveness, and efficient charge-transfer interactions. These findings underscore the promise of $CrTe_2$-based nanosensors in next-generation environmental and industrial sensing technologies.

## Acknowledgments

Authors acknowledge funding from the Australian Government through the ARC Discovery Project (DP250101156). The Centre for Advanced Computation supports this work at the Korea Institute for Advanced Study. PP gratefully acknowledges financial support from CENCON.

## References

(1) Tai, A. P. K.; Martin, M. V.; Heald, C. L. Threat to Future Global Food Security from Climate Change and Ozone Air Pollution. *Nature Clim Change* **2014**, *4* (9), 817–821. https://doi.org/10.1038/nclimate2317.

(2) Hulin, M.; Simoni, M.; Viegi, G.; Annesi-Maesano, I. Respiratory Health and Indoor Air Pollutants Based on Quantitative Exposure Assessments. *European Respiratory Journal* **2012**, *40* (4), 1033–1045. https://doi.org/10.1183/09031936.00159011.

(3) Alcheikhhamdon, Y.; Hoorfar, M. Natural Gas Purification from Acid Gases Using Membranes: A Review of the History, Features, Techno-Commercial Challenges, and Process Intensification of Commercial Membranes. *Chemical Engineering and Processing - Process Intensification* **2017**, *120*, 105–113. https://doi.org/10.1016/j.cep.2017.07.009.

(4) Chen, Z.; Sidell, M. A.; Huang, B. Z.; Chow, T.; Eckel, S. P.; Martinez, M. P.; Gheissari, R.; Lurmann, F.; Thomas, D. C.; Gilliland, F. D.; Xiang, A. H. Ambient Air Pollutant Exposures and COVID-19 Severity and Mortality in a Cohort of Patients with COVID-19 in Southern California. *Am J Respir Crit Care Med* **2022**, *206* (4), 440–448. https://doi.org/10.1164/rccm.202108-1909OC.

(5)  Zhu, Y.; Chen, G. Simulation and Assessment of SO2 Toxic Environment after Ignition of Uncontrolled Sour Gas Flow of Well Blowout in Hills. *Journal of Hazardous Materials* **2010**, *178* (1), 144–151. https://doi.org/10.1016/j.jhazmat.2010.01.055.

(6)  Oliverio, S.; Varlet, V. New Strategy for Carbon Monoxide Poisoning Diagnosis: Carboxyhemoglobin (COHb) vs Total Blood Carbon Monoxide (TBCO). *Forensic Science International* **2020**, *306*, 110063. https://doi.org/10.1016/j.forsciint.2019.110063.

(7)  Yong, Y.; Cui, H.; Zhou, Q.; Su, X.; Kuang, Y.; Li, X. C2N Monolayer as NH3 and NO Sensors: A DFT Study. *Applied Surface Science* **2019**, *487*, 488–495. https://doi.org/10.1016/j.apsusc.2019.05.040.

(8)  Park, J.-H.; Cho, J. H.; Kim, Y. J.; Kim, E. S.; Han, H. S.; Shin, C.-H. Hydrothermal Stability of Pd/ZrO2 Catalysts for High Temperature Methane Combustion. *Applied Catalysis B: Environmental* **2014**, *160–161*, 135–143. https://doi.org/10.1016/j.apcatb.2014.05.013.

(9)  Woo, H.-S.; Na, C. W.; Lee, J.-H. Design of Highly Selective Gas Sensors via Physicochemical Modification of Oxide Nanowires: Overview. *Sensors* **2016**, *16* (9), 1531. https://doi.org/10.3390/s16091531.

(10) Tang, Y.; Chen, W.; Li, C.; Pan, L.; Dai, X.; Ma, D. Adsorption Behavior of Co Anchored on Graphene Sheets toward NO, SO2, NH3, CO and HCN Molecules. *Applied Surface Science* **2015**, *342*, 191–199. https://doi.org/10.1016/j.apsusc.2015.03.056.

(11) Zhan, D.; Yan, J. X.; Ni, Z. H.; Sun, L.; Lai, L. F.; Liu, L.; Liu, X. Y.; Shen, Z. X. Bandgap-Opened Bilayer Graphene Approached by Asymmetrical Intercalation of Trilayer Graphene. *Small* **2015**, *11* (9–10), 1177–1182. https://doi.org/10.1002/smll.201402728.

(12) Chhowalla, M.; Shin, H. S.; Eda, G.; Li, L.-J.; Loh, K. P.; Zhang, H. The Chemistry of Two-Dimensional Layered Transition Metal Dichalcogenide Nanosheets. *Nature Chem* **2013**, *5* (4), 263–275. https://doi.org/10.1038/nchem.1589.

(13) Zeng, Y.; Lin, S.; Gu, D.; Li, X. Two-Dimensional Nanomaterials for Gas Sensing Applications: The Role of Theoretical Calculations. *Nanomaterials* **2018**, *8* (10), 851. https://doi.org/10.3390/nano8100851.

(14) Zhang, X.; Teng, S. Y.; Loy, A. C. M.; How, B. S.; Leong, W. D.; Tao, X. Transition Metal Dichalcogenides for the Application of Pollution Reduction: A Review. *Nanomaterials* **2020**, *10* (6), 1012. https://doi.org/10.3390/nano10061012.

(15) Wu, Y.; Joshi, N.; Zhao, S.; Long, H.; Zhou, L.; Ma, G.; Peng, B.; Oliveira Jr, O. N.; Zettl, A.; Lin, L. NO2 Gas Sensors Based on CVD Tungsten Diselenide Monolayer. *Applied Surface Science* **2020**, *529*, 147110. https://doi.org/10.1016/j.apsusc.2020.147110.

(16) Cui, H.; Zhang, X.; Zhang, J.; Zhang, Y. Nanomaterials-Based Gas Sensors of SF6 Decomposed Species for Evaluating the Operation Status of High-Voltage Insulation Devices. *High Voltage* **2019**, *4* (4), 242–258. https://doi.org/10.1049/hve.2019.0130.

(17) Cao, J.; Zhou, J.; Chen, J.; Wang, W.; Zhang, Y.; Liu, X. Effects of Phase Selection on Gas-Sensing Performance of MoS2 and WS2 Substrates. *ACS Omega* **2020**, *5* (44), 28823–28830. https://doi.org/10.1021/acsomega.0c04176.

(18) Panigrahi, P.; Hussain, T.; Karton, A.; Ahuja, R. Elemental Substitution of Two-Dimensional Transition Metal Dichalcogenides (MoSe2 and MoTe2): Implications for Enhanced Gas Sensing. *ACS Sens.* **2019**, *4* (10), 2646–2653. https://doi.org/10.1021/acssensors.9b01044.

(19) Wu, P.; Huang, M. Mechanism of Adsorption and Gas-Sensing of Hazardous Gases by MoS2 Monolayer Decorated by Pdn (*n*=1–4) Clusters. *Colloids and Surfaces A: Physicochemical and Engineering Aspects* **2024**, *695*, 134200. https://doi.org/10.1016/j.colsurfa.2024.134200.

(20) Panigrahi, P.; Srinivasan, U.; Sharma, M.; Bae, H.; Lee, H.; Panigrahi, A.; Hussain, T. Detecting Specific Volatile Organic Compounds in Aquaculture Monitoring Using Tungsten Diselenide Monolayers. *Materials Today Chemistry* **2025**, *45*, 102656. https://doi.org/10.1016/j.mtchem.2025.102656.

(21) Kaewmaraya, T.; Amorim, R. G.; Thatsami, N.; Moontragoon, P.; Pinitsoontorn, S.; Bae, H.; Lee, H.; Nasiri, N.; Hussain, T. Highly Efficient Room-Temperature Ethylene Sensing with Molybdenum Based Transition Metal Dichalcogenides. *Applied Surface Science* **2025**, *697*, 162984. https://doi.org/10.1016/j.apsusc.2025.162984.

(22) Panigrahi, P.; Kotmool, K.; Khammuang, S.; Bae, H.; Gulati, V.; Hussain, T. Smart Sensing Characteristics of Tungsten Diselenide (WSe2) Monolayers toward Depression-Related Volatile Organic Compounds. *ACS Appl. Nano Mater.* **2025**, *8* (11), 5685–5693. https://doi.org/10.1021/acsanm.5c00211.

(23) Mohammadzadeh, M. R.; Hasani, A.; Hussain, T.; Ghanbari, H.; Fawzy, M.; Abnavi, A.; Ahmadi, R.; Kabir, F.; De Silva, T.; Rajapakse, R. K. N. D.; Adachi, M. M. Enhanced Sensitivity in Photovoltaic 2D MoS2/Te Heterojunction VOC Sensors. *Small* **2024**, *20* (49), 2402464. https://doi.org/10.1002/smll.202402464.

(24) Panigrahi, P.; Pal, Y.; Kaewmaraya, T.; Bae, H.; Nasiri, N.; Hussain, T. Molybdenum Carbide MXenes as Efficient Nanosensors toward Selected Chemical Warfare Agents. *ACS Appl. Nano Mater.* **2023**, *6* (10), 8404–8415. https://doi.org/10.1021/acsanm.3c00686.

(25) Kadam, S. A. Advancements in Monolayer TMD-Based Gas Sensors: Synthesis, Mechanisms, Electronic Structure Engineering, and Flexible Wearable Sensors for Real-World Applications and Future Prospects. *Chemical Engineering Journal* **2025**, *517*, 164223. https://doi.org/10.1016/j.cej.2025.164223.

(26) Mirzaei, A.; Kim, J.-Y.; Kim, H. W.; Kim, S. S. Resistive Gas Sensors Based on 2D TMDs and MXenes. *Acc. Chem. Res.* **2024**, *57* (16), 2395–2413. https://doi.org/10.1021/acs.accounts.4c00323.

(27) Anisha; Singh, M.; Kumar, R.; Srivastava, S.; Tankeshwar, K. Tuning of Thermoelectric Performance of CrSe2 Material Using Dimension Engineering. *Journal of Physics and Chemistry of Solids* **2023**, *172*, 111083. https://doi.org/10.1016/j.jpcs.2022.111083.

(28) Li, B.; Wan, Z.; Wang, C.; Chen, P.; Huang, B.; Cheng, X.; Qian, Q.; Li, J.; Zhang, Z.; Sun, G.; Zhao, B.; Ma, H.; Wu, R.; Wei, Z.; Liu, Y.; Liao, L.; Ye, Y.; Huang, Y.; Xu, X.; Duan, X.; Ji, W.; Duan, X. Van Der Waals Epitaxial Growth of Air-Stable CrSe2 Nanosheets with Thickness-Tunable Magnetic Order. *Nat. Mater.* **2021**, *20* (6), 818–825. https://doi.org/10.1038/s41563-021-00927-2.

(29) Lin, L.; Sun, Y.; Xie, K.; Shi, P.; Yang, X.; Wang, D. First-Principles Study on the Catalytic Performance of Transition Metal Atom-Doped CrSe2 for the Oxygen Reduction Reaction. *Phys. Chem. Chem. Phys.* **2023**, *25* (22), 15441–15451. https://doi.org/10.1039/D3CP00223C.

(30) Katanin, A. A.; Agapov, E. M. Magnetic Properties of Monolayer, Multilayer, and Bulk ${\mathrm{CrTe}}_{2}$. *Phys. Rev. B* **2025**, *111* (3), 035118. https://doi.org/10.1103/PhysRevB.111.035118.

(31) Zhang, X.; Lu, Q.; Liu, W.; Niu, W.; Sun, J.; Cook, J.; Vaninger, M.; Miceli, P. F.; Singh, D. J.; Lian, S.-W.; Chang, T.-R.; He, X.; Du, J.; He, L.; Zhang, R.; Bian, G.; Xu, Y. Room-Temperature Intrinsic Ferromagnetism in Epitaxial CrTe2 Ultrathin Films. *Nat Commun* **2021**, *12* (1), 2492. https://doi.org/10.1038/s41467-021-22777-x.

(32) Zhu, W.; Wang, P.; Xie, K.; Zhang, C.; Dong, Z.; Lin, L. First-Principles Study of the Adsorption and Sensing Properties of Transition Metal-Modified CrSe2 for CH4, H2S, and CO. *Colloids and Surfaces A: Physicochemical and Engineering Aspects* **2025**, *708*, 136006. https://doi.org/10.1016/j.colsurfa.2024.136006.

(33) Hafner, J. Ab-Initio Simulations of Materials Using VASP: Density-Functional Theory and Beyond. *Journal of Computational Chemistry* **2008**, *29* (13), 2044–2078. https://doi.org/10.1002/jcc.21057.

(34) Kresse, G.; Joubert, D. From Ultrasoft Pseudopotentials to the Projector Augmented-Wave Method. *Phys. Rev. B* **1999**, *59* (3), 1758–1775. https://doi.org/10.1103/PhysRevB.59.1758.

(35) Kresse, G.; Furthmüller, J. Efficiency of Ab-Initio Total Energy Calculations for Metals and Semiconductors Using a Plane-Wave Basis Set. *Computational Materials Science* **1996**, *6* (1), 15–50. https://doi.org/10.1016/0927-0256(96)00008-0.

(36) Perdew, J. P.; Burke, K.; Ernzerhof, M. Generalized Gradient Approximation Made Simple. *Phys. Rev. Lett.* **1996**, *77* (18), 3865–3868. https://doi.org/10.1103/PhysRevLett.77.3865.

(37) Perdew, J. P.; Chevary, J. A.; Vosko, S. H.; Jackson, K. A.; Pederson, M. R.; Singh, D. J.; Fiolhais, C. Atoms, Molecules, Solids, and Surfaces: Applications of the Generalized Gradient Approximation for Exchange and Correlation. *Phys. Rev. B* **1992**, *46* (11), 6671–6687. https://doi.org/10.1103/PhysRevB.46.6671.

(38) Blöchl, P. E. Projector Augmented-Wave Method. *Phys. Rev. B* **1994**, *50* (24), 17953–17979. https://doi.org/10.1103/PhysRevB.50.17953.

(39) Grimme, S.; Antony, J.; Ehrlich, S.; Krieg, H. A Consistent and Accurate Ab Initio Parametrization of Density Functional Dispersion Correction (DFT-D) for the 94 Elements H-Pu. *J. Chem. Phys.* **2010**, *132* (15), 154104. https://doi.org/10.1063/1.3382344.

(40) Zhang, Y.-H.; Chen, Y.-B.; Zhou, K.-G.; Liu, C.-H.; Zeng, J.; Zhang, H.-L.; Peng, Y. Improving Gas Sensing Properties of Graphene by Introducing Dopants and Defects: A First-Principles Study. *Nanotechnology* **2009**, *20* (18), 185504. https://doi.org/10.1088/0957-4484/20/18/185504.

(41) Yu, M.; Trinkle, D. R. Accurate and Efficient Algorithm for Bader Charge Integration. *J. Chem. Phys.* **2011**, *134* (6), 064111. https://doi.org/10.1063/1.3553716.

(42) Sharma, S. B.; Paudel, R.; Adhikari, R.; Kaphle, G. C.; Paudyal, D. Structural Deformation and Mechanical Response of CrS2<math><msub Is="true"><mrow

Is="true"></Mrow><mrow Is="true"><mn Is="true">2</Mn></Mrow></Msub></Math>, CrSe2<math><msub Is="true"><mrow Is="true"></Mrow><mrow Is="true"><mn Is="true">2</Mn></Mrow></Msub></Math> and Janus CrSSe. *Physica E: Low-dimensional Systems and Nanostructures* **2023**, *146*, 115517. https://doi.org/10.1016/j.physe.2022.115517.

(43) Liu, Y.; Kwon, S.; de Coster, G. J.; Lake, R. K.; Neupane, M. R. Structural, Electronic, and Magnetic Properties of $ {\mathrm{CrTe}}_{2}$. *Phys. Rev. Mater.* **2022**, *6* (8), 084004. https://doi.org/10.1103/PhysRevMaterials.6.084004.

(44) Pourebrahimi, S.; Pirooz, M.; Kazemeini, M.; Vafajoo, L. Synthesis, Characterization, and Gas (SO2, CO2, NO2, CH4, CO, NO, and N2) Adsorption Properties of the CTF-1 Covalent Triazine Framework-Based Porous Polymer: Experimental and DFT Studies. *J Porous Mater* **2024**, *31* (2), 643–657. https://doi.org/10.1007/s10934-023-01538-9.

(45) Doust Mohammadi, M.; Louis, H.; Chukwu, U. G.; Bhowmick, S.; Rasaki, M. E.; Biskos, G. Gas-Phase Interaction of CO, CO2, H2S, NH3, NO, NO2, and SO2 with Zn12O12 and Zn24 Atomic Clusters. *ACS Omega* **2023**, *8* (23), 20621–20633. https://doi.org/10.1021/acsomega.3c01177.

(46) Lin, L.; Feng, Z.; Dong, Z.; Hu, C.; Han, L.; Tao, H. DFT Study on the Adsorption of CO, NO2, SO2 and NH3 by Te Vacancy and Metal Atom Doped MoTe2 Monolayers. *Physica E: Low-dimensional Systems and Nanostructures* **2023**, *145*, 115489. https://doi.org/10.1016/j.physe.2022.115489.